\documentclass[prd,onecolumn]{revtex4}
\usepackage{dcolumn}
\usepackage{multirow}
\usepackage{graphicx}
\usepackage{amssymb}
\usepackage{bm}
\usepackage{hyperref}
\usepackage{epstopdf}
\usepackage{color}
\usepackage{mathrsfs}
\usepackage{amsmath,amssymb,amsthm}
\begin{document}
\title{Wormholes in a viscous universe}
\author{Deng Wang}
\email{cstar@nao.cas.cn}
\affiliation{National Astronomical Observatories, Chinese Academy of Sciences, Beijing, 100012, China}

\begin{abstract}
We study the static and spherically symmetric wormhole spacetime configurations at present time in a viscous universe. Considering three classes of viscous models, i.e., bulk viscosity as a function of Hubble constant $H_0$, present cosmic temperature $T_0$ and dark energy density $\rho_0$, respectively, we obtain several wormhole solutions. Through the analysis for the anisotropic solutions, we conclude that, to some extent, these three classes of viscous models have very high degeneracy with each other, and deviate slightly from the ideal fluid consequence. Subsequently, without the loss of generality, to investigate the traversabilities, energy conditions and stability for the wormhole solution, we study the wormhole solution of the constant redshift function of the viscous $\omega$CDM model with a constant bulk viscosity coefficient. We obtain the following conclusions: the value of traversal velocity decreases for decreasing bulk viscosity, and the traversal velocity for a traveler depends on not only the wormhole geometry but also the effects of cosmological background evolution; the null energy condition will be violated more clearly when the bulk viscosity or the cosmic expansion velocity decreases; for the case of a positive surface energy density, the range of the junction radius decreases and the values of the parameter $\lambda$ are further restricted when the throat radius of the wormhole $r_0/M$ increases, bulk viscosity coefficient $\zeta_0$ decreases and present-day Hubble parameter $H_0$ decreases, respectively; for the case of a negative surface energy density, by increasing $r_0/M$, decreasing $\zeta_0$ and increasing $H_0$, the values of the parameter $\lambda$ are less restricted, one may conclude that the total stability region increases, and the range of the junction radius increases, increases and decreases, respectively.
\end{abstract}
\maketitle
\section{introduction}
In recent years, the cosmological observations has provided increasing evidence that our universe is under a phase of accelerated expansion \cite{1,2}. In order to explain the acceleration expansion phenomena, cosmologists during the past few decades introduce a new fluid, which possesses a negative enough pressure, named dark energy. According to the recent observational evidence, especially from the Type Ia supernovae project, the Sloan Digital Sky Survey (SDSS) project, the Wilkinson Microwave Anisotropy Probe (WMAP) satellite missions, the Planck satellite missions and galaxies clustering, we live in a favored spatially flat universe consisting approximately of 4\% baryonic matter, 23\% dark matter and 73\% dark energy. The simplest candidate for dark energy is the cosmological constant scenario, $\Lambda$, first introduced by Einstein in order to keep the universe static, but dismissed soon when the universe is found to be undergoing an accelerated phase. So far, this mysterious constant has been reintroduced several times in history in attempts to explain several astrophysical phenomena, including most recently the flat spatial geometry implied by the Planck-2015 \cite{3}. A cosmological constant is described by a single parameter, the inclusion of which brings the $\Lambda$CDM ($\Lambda$-Cold Dark Matter) model into excellent agreement with the current observations. However, there exist two fatal problems in this model \cite{4}. The first one is the fine-tuning problem: theoretical estimates for the vacuum density are many orders of magnitude larger than its observed value, i.e., the famous 120-orders-of-magnitude discrepancy that makes the vacuum explanation suspicious. The second one is the coincidence problem, i.e., $\Omega_{\Lambda}$ and $\Omega_{m}$ are of the same order of magnitude only at present, which indicates our epoch as a very special time in the evolution of the universe. Besides, a positive cosmological constant is inconsistent with perturbed string theory \cite{5}. Therefore, the actual origin of the dark energy might not be the $\Lambda$. This lack of a clear theoretical understanding has motivated the development of a wide variety of alternative models to describe the universe with dark energy.

In the present paper, we shall pay our attention to study the astrophysical scales properties in the universe by assuming the dark energy fluid is permeated everywhere rather than concentrating on studying the evolution of the universe.  Moreover, in our entire universe, the existence of wormholes is a substantially appealing problem in physics both at micro and macro scales. There is no doubt that together with black holes, pulsars, quasars and white dwarfs, etc., wormholes constituting the most intriguing and puzzling celestial bodies may provide a new window for new physics. Thus, it is very constructive necessary to make a brief review about the developments of wormholes.

A wormhole is a feature of space that is essentially a ``shortcut'' from one point in the universe to another point in the universe, allowing travel between them that is faster than it would take light to make the journey through the normal space. Thus, wormholes are tunnels or handles in the spacetime topology that connects different universes or widely separated regions of our universe via a throat. The earliest significant contribution we are aware of is the introduction, in 1935, of the object now referred to as Einstein-Rosen bridge \cite{6}. This field then lay dormant for twenty years until Wheeler first coined the term `` wormhole '' and introduced his idea of`` spacetime foam '' \cite{3}. A thirty year interregnum followed, punctuated by isolated contributions, until the major revival of interest following the 1988 paper by Morris and Thorne \cite{8}. In their elegant work, a new class of solutions of the Einstein field equations was presented, which describe wormholes that, in principle, could be traversable by human beings. It is essential in these solutions that the wormhole possesses a throat at which there is no horizon. In the meanwhile, this property, together with the field equations, places an extreme constraint on the material that generates the wormhole spacetime curvature. This leads to such material must violate all the energy conditions (null, weak, strong and dominant) that underlie some deeply cherished theorems in general relativity. Nonetheless, we could not easily rule out the existence of the material; and quantum field theory gives tantalizing hints that such material might, in fact, be possible. More specifically, they prove black holes are unstable for interstellar travel and Schwarzschild wormholes is non-traversable, analyze the properties of traversable wormholes, calculate out the mathematical details of traversable wormholes and explore the stability of the wormhole.
Subsequently, Visser \cite{9} and Possion and Visser \cite{10} conjectured that all `` exotic '' matter is confined to a thin shell between universes. Their model is constructed surgically by grafting two Schwarzschild spacetimes together so that no horizon exists. It is so-called thin shell wormholes which has been widely investigated by many authors in various alternative cosmological models and theories of gravity. Furthermore, Visser et al. develops the notion of `` volume integral quantifier '' (VIQ), and one may obtain the total amount of energy condition violating matter by calculating the VIQ \cite{11}. After that, the subject has grown substantially, and it is now almost out of control and there are many branches out of the special space-time structures etc., such as energy conditions, theoretical construction, stability, time machines, and astrophysical signatures. A quarter century later, a follow-up is written by Thorne which has two purposes \cite{12}:

$\star$ to explain how Interstellar's wormhole images were constructed and explain the decisions made on the way to their final form.

$\star$ to present this explanation that may be useful to students and teachers in elementary courses on general relativity.

We could find that the follow-up emphasize mainly the feasibility and the real status of wormholes in our real universe.

Recently, because of the new discovery that our universe is undergoing an accelerated expansion, an increasing interest to the subjects has arisen significantly in connection with the discovery. Since in both cases the null energy condition (NEC), $T_{\mu\nu}k^{\mu}k^{\nu}>0$, is violated and consequently all of other energy conditions, where $T_{\mu\nu}$ is the stress-energy tensor and $k^{\mu}$ any future directed null vector. Therefore, an interesting and surprising overlap between two seemingly separated subjects occurs. A simple way to parameterize the dark energy is by an equation of state of this form $-K=p/\rho$, where $p$ is the spatially homogeneous pressure and $\rho$ the energy density of the dark energy. A value of $K>1/3$ is required for cosmic expansion, and $K=1$ a cosmological constant. The particular case $K=2/3$ is extensively analyzed in \cite{13}. In quintessence models, the parameter range is $1/3<K<1$ and the dark energy decreases with a scale factor $a(t)$ as $\rho \propto a^{3(K-1)}$. In phantom models, the the parameter range is $K>1$ and we can also find that the NEC is naturally violated, $p+\rho<0$, in the phantom case.

In the literature, all the authors almost construct wormholes by only using the linear equation of state, namely, assuming that the universe is filled with a perfect fluid. However, the dynamical behaviors of the true universe are physically dominated by a non-ideal cosmological fluid. Based on this concern, in our previous works \cite{14,15}, we have investigated the wormhole spacetime configurations in a non-ideal fluid universe by using the Shan-Chen equation of state. Actually, it is worth noting that the the Shan-Chen cosmology is essentially equivalent to the viscous cosmology. In the present paper, we are dedicated to investigate the wormholes in the bulk viscous cosmology. Two authors have studied some aspects of wormholes in the bulk viscosity cosmology: Hristu Culetu develops a classical membrane model for a dynamical wormhole \cite{16}. This model leads to an expanding throat and a negative bulk viscosity coefficient $-1/{6\pi}$ on the throat, implying the instability against perturbations; Mubasher Jamil has investigated the effects of the accretion of phantom energy with non-zero bulk viscosity onto a Morris-Thorne wormhole, and find that if the bulk viscosity is enough large, the mass of wormhole increases rapidly as compared to small or zero bulk viscosity \cite{17}. Furthermore, we are aiming at obtaining the clear expressions of the shape function and the redshift function in the solutions of Einstein field equations, and study the related properties and characteristics of the obtained wormhole spacetime structure. More precisely, we explore the wormhole configurations in different viscous models starting from the bulk viscosity equation of state. From this, we should review the history of bulk viscosity and its applications in modern cosmology and theories of gravity.

We have felt an increasing interest and attention in viscous cosmology during the past several decades. This contrasts the traditional approach, in which the cosmic fluid has usually be taken to be perfect (non-viscous). From a hydrogynamicist's point of view this is to some extent surprising, since there are several situations in fluid mechanics, even in homogeneous space without boundaries, where the inclusion of viscosity becomes mandatory. Especially, this is so in connection with turbulence phenomena. In history, perfect fluid models have for a long time been used in cosmological theory and the introduction of viscosity concepts came later. Misner was probably the first to introduce viscosity in cosmological theory in connection with his study of how initial anisotropies in the early universe become relaxed \cite{18}. Gr{\o}n has given an extensive review of the subject, the reader is referred to it for detailed information \cite{19}. Recently, Dou and Meng, Ma and Meng, Nojiri and Brevik give new different reviews from different points of view, the reader is referred to them for more details \cite{20,21,22,23}.

As is well known, the perfect fluid assumption is assertive since it implies no dissipation, which actually exists widely and intuitively play a crucial role in the evolution of the universe, especially in the very early hot stages. To be more realistic, models of imperfect fluid are invoked by introducing viscosity into investigation. More precisely, to the first order in deviation from thermal equilibrium there are two such coefficients, the shear viscosity $\xi$, most often being the dominant one, and the bulk viscosity $\zeta$ \cite{24}. The simplest theory of this kind is the case of a constant bulk viscosity, regarded as a equivalence to the cosmological constant scenario. A widely investigated case is that bulk viscosity as a function of Hubble parameter, which has been proved to be well consistent with observed late-time acceleration and recover both the matter and dark energy dominant eras. Another intriguing case is that bulk viscosity behaves as a parameterized power-law with respect to the matter density, which could be shown to be similar to the Chaplygin gas model. In addition, temperature-dependent viscosity from classical statistical physics is introduced into observational cosmology and  behaves well as compared with the observed data. Furthermore, viscosity term containing high derivatives of scale factor and redshift is also considered as a more general theory \cite{25,26,27,28,29,30}. In the past few years, viscosity-induced crossing of the phantom barrier in Einstein gravity and modified gravity and various models of dark energy coupled with dark matter in a dissipative universe is studied. Especially in some special cases, the Little Rip, Pseudo Rip and bounce cosmological models can be related to each other via the bulk viscosity concept. Moreover, it is necessary to note that viscous Ricci dark energy (RDE) is also considered in the literature \cite{31}. In the RDE model without bulk viscosity, the universe is younger than some old objects at some redshifts. Since the age of the universe should be longer than any objects in the universe, the RDE model suffers the age problem, especially when we consider the object APM $08279 + 5255$ at $z = 3.91$, whose age is $t = 2.1$ Gyr, $z$ being the redshift. Nevertheless, once the bulk viscosity effect is taken into account, this age problem is alleviated. In this paper, we focus on the aspects of wormhole physics of three classes of models in the bulk viscous cosmology: bulk viscosity as a function of Hubble parameter, temperature and matter density, respectively. In what follows, it is necessary to make a brief review for the basic formalism of the viscous cosmology.

We use the convention in which the Minkowski metric is $(-1, 1, 1, 1)$. Greek indices are summed from $0$ to $3$, Latin indices are summed from 1 to 3. In the scheme of imperfect fluids, the energy stress tensor can be written as in the following manner,
\begin{equation}
T_{\mu\nu}=\rho U_{\mu}U_{\nu}+(p-\theta\zeta)h_{\mu\nu}-2\xi\sigma_{\mu\nu}+Q_{\mu}U_{\nu}+Q_{\nu}U_{\mu}, \label{1}
\end{equation}
where $\rho$ is the mass-energy density, p the pressure, $U_{\mu}=(1, 0, 0, 0)$ the four-velocity of the cosmic fluid in comoving coordinates, $h_{\mu\nu}=g_{\mu\nu}+ U_{\mu}U_{\nu}$ the projection tensor, $\omega_{\mu\nu}=h_{\mu}^{\iota}h_{\nu}^{\upsilon}U_{[\iota;\upsilon]}=\frac{1}{2}(U_{\mu;\iota}h_{\nu}^{\iota}-U_{\nu;\iota}h_{\mu}^{\iota})$ the rotation tensor, $\theta_{\mu\nu}=h_{\mu}^{\iota}h_{\nu}^{\upsilon}U_{(\iota;\upsilon)}=\frac{1}{2}(U_{\mu;\iota}h_{\nu}^{\iota}+U_{\nu;\iota}h_{\mu}^{\iota})$ is the expansion tensor, $\theta=\theta_{\mu}^{\nu}=U_{;\mu}^{\mu}$ the expansion scalar, $\sigma_{\mu\nu}=\theta_{\mu}^{\nu}-\frac{1}{3}h_{\mu}^{\nu}\theta$, $\zeta$ the bulk viscosity, $\xi$ the shear viscosity, $U_{\mu;\nu}=\omega_{\mu\nu}+\sigma_{\mu\nu}+\frac{1}{3}h_{\mu}^{\nu}\theta-A_{\mu}U_{\nu}$ the decomposition of the covariant derivative of the fluid velocity, $A_{\nu}=\dot{U}_{\mu}=U_{\nu}U_{\mu;\nu}$ the four-acceleration of the fluid, and $Q^{\mu}=-\kappa h^{\mu\nu}(T_{,\nu}+TA_{\nu})$ the heat flux density four-vector with k the thermal conductivity.

In the case of thermal equilibrium, $Q_{\mu}=0$, additionally, the term of shear viscosity vanished when a completely isotropic universe is assumed. In the isotropic and homogeneous Friedmann-Robertson-Walker (FRW) metric, Eq.(1) can be rewritten as
\begin{equation}
T_{\mu\nu}=\rho U_{\mu}U_{\nu}+(p-\theta\zeta)h_{\mu\nu}, \label{2}
\end{equation}
where $\theta=3H=3\frac{\dot{a}}{a}$ and the dot denotes differential with respect to the cosmic time $t$.

In general, wormholes can be classified into two categories-Euclidean wormholes and Lorentzian wormholes based on the famous Gorech theorem. The Euclidean wormholes comes from Euclidean quantum gravity and the Lorentzian wormholes that are static spherically symmetric solutions of Einstein field equations. Here we just explore the Lorentzian wormholes in a viscous universe. It is worth noting that our work is mainly motivated by three earlier studies \cite{32,33,34} that have demonstrated the possible existence of wormholes in the outer regions of the galactic halo and in the central parts of the halo, respectively, based on NFW (Navarro-Frenk-White) density profile and the URC (Universal Rotation Curve) dark matter model \cite{35}. Especially, the second result is an important compliment to the earlier result, thereby confirming the possible existence of wormholes in most of the spiral galaxies. After their interesting and suggestive works, immediately, A. \"{O}vg\"{u}n et al. considered the existence of wormholes in the spherical stellar systems. In addition, it is shown that based on the Einasto model \cite{36,37,38} wormholes in the outer regions of spiral galaxies are possible while the central regions prohibit such space-time structure formations.

The rest of this paper is outlined in the following manner. In the next section, we make a review of mathematical details about wormhole calculations and assume the simplest anisotropic pressure for cosmic fluid. In Section $3$, there are three classes of viscous models to be discussed and obtain several solutions in which, for every model, we solve the Einstein equations by both inserting the shape function $b(r)=r_0+\frac{1}{K}(r-r_0)$ and the redshift function $\Phi=C$: in Subsection $1$, we investigate the wormhole solutions where the bulk viscosity as a function of Hubble parameter $H$, and consider the relatively simple linear model, not involving the quadratic term $H^2$, and power law model; in Subsection $2$, we consider the bulk viscosity as a function of temperature $T$ and analyze the power law model and obtain two solutions; in Subsection $3$, we study the viscosity term as a function of the dark energy density $\rho$, similarly, explore the power law model and obtain two solutions. In Sections $4-6$, as a concrete case, we investigate the traversabilities, energy conditions and stability for the wormhole solution of the constant redshift function of the viscous $\omega$CDM model with a constant bulk viscosity coefficient. In the final section, we discuss and conclude (we use units $G=c=1$).

\section{Wormholes as special space-time structures}
In the present study, we consider a spacetime metric for the wormholes in a static spherically symmetric form
\begin{equation}
ds^2=-Udt^2+\frac{dr^2}{V}+r^2d\Omega^2, \label{3}
\end{equation}
and U and V can be defined as
\begin{equation}
V=1-\frac{b(r)}{r}, \quad U=e^{2\Phi(r)}, \label{4}
\end{equation}
where $r$ is the radial coordinate which runs in the range $r_0\leq r<\infty$ ($r_0$ is the throat radius of the wormhole), and $d\Omega^2$ is the unit line element on the sphere.  To form a wormhole, the redshift function $\Phi(r)$ should be finite and non-vanishing in order to avoid an event horizon in the vicinity of $r_0$, and the shape function $b(r)$, which determines the spatial shape of the wormhole when viewed, must obey the usual flare-out conditions at the throat
\begin{equation}
b(r_0)=r_0, \label{5}
\end{equation}
\begin{equation}
b'(r_0)<1, \label{6}
\end{equation}
\begin{equation}
b(r)<r,r>r_0. \label{7}
\end{equation}
If one expects to obtain a traversable wormhole, the asymptotically flatness condition is required, that is, $b(r)/r\rightarrow0$ as $r\rightarrow\infty$.
Utilizing the Einstein field equation, $G_{\alpha\beta}=8\pi T_{\alpha\beta}$, in an orthonormal reference frame, we have the following equations
\begin{equation}
\frac{b'}{8\pi r^2}=\rho(r), \label{8}
\end{equation}
\begin{equation}
\frac{U'}{U}=\frac{8\pi p_r r^3+b}{r(r-b)}, \label{9}
\end{equation}
where $T_{t}^t=-\rho, T_{r}^r=p_r, T_{\nu}^\mu$ is the stress-energy tensor.

One can also derive from the conservation law of the stress-energy tensor $T^\nu_{\mu;\nu} = 0$ with $\mu=r$, which can be interpreted as the hydrostatic equation for equilibrium for the material threading the wormhole, that
\begin{equation}
p_\perp=\frac{r}{2}[p'_r+\frac{2p_r}{r}+\frac{U'}{2U}(p_r+\rho)], \label{10}
\end{equation}
where $T_\theta^\theta=T_\phi^\phi=p_\perp$. In addition, we assume the pressures are anisotropic and just consider the simplest case between the tangential pressure and the energy density but with $p_{\perp}\neq p_r$,
\begin{equation}
p_\perp=\alpha\rho. \label{11}
\end{equation}
\section{models}
In the following context, considering the radial pressure $p_r=-K\rho-\theta\zeta$ ($\theta=3H_0$) at thermodynamical equilibrium  and the equation of state parameter $K>1$, we investigate three classes of viscous models, i.e., bulk viscosity as a function of Hubble parameter $H_0$, temperature $T$ and dark energy density $\rho$, respectively, and obtain several wormhole solutions.

\subsection{bulk viscosity as a function of Hubble constant: $\zeta=\zeta(H_0)$}
\subsubsection{linear model: $\zeta=\zeta_0+\zeta_1H_0$}
For the linear models, we just consider the relatively simple case, where the bulk viscosity can be expressed as
\begin{equation}
\zeta=\zeta_0+\zeta_1H_0 \label{12}
\end{equation}
where $\zeta_0$ and $\zeta_1$ are two viscosity parameters conventionally. Hence, the radial pressure is
\begin{equation}
p_r=-K\rho-3\zeta_0H_0-3\zeta_1H_0^2. \label{13}
\end{equation}
It is easy to see that the bulk viscosity coefficient reduces to be a constant when $\zeta_1=0$.

Here we consider two specific shape functions that are widely investigated in the literature \cite{14,15,39}. At first, we take the shape function $b(r)=r_0+\frac{1}{K}(r-r_0)$, then Eqs. (\ref{8}-\ref{9}) become
\begin{equation}
\rho(r)=\frac{1}{8K\pi r^2}, \label{14}
\end{equation}
\begin{equation}
\frac{U'}{U}=-\frac{1}{r}-\frac{24\pi H_0(\zeta_0+\zeta_1H_0)r^2}{(r-r_0)(1-\frac{1}{K})}. \label{15}
\end{equation}
It follows that
\begin{equation}
U(r)=\frac{A}{r}e^{\frac{12H_0K\pi(\zeta_0+\zeta_1H_0)[r(2r_0+r)+2r_0^2\ln(r-r_0)]}{1-K}}. \label{16}
\end{equation}
Thus, the metric becomes
\begin{equation}
ds^2=-\frac{A}{r}e^{\frac{12H_0K\pi(\zeta_0+\zeta_1H_0)[r(2r_0+r)+2r_0^2\ln(r-r_0)]}{1-K}}dt^2+\frac{dr^2}{(1-\frac{1}{K})(r-r_0)}+r^2d\Omega^2, \label{17}
\end{equation}
where $A$ is an arbitrary constant that affects the normalization of time.

Then the next step is to calculate out the $\alpha$ in Eq. (\ref{11}). Substituting Eq. (\ref{11}) and Eq. (\ref{13}) and Eqs. (\ref{14}-\ref{15}) into Eq. (\ref{10}), we can obtain
\begin{equation}
\alpha=\alpha(r)=\frac{K-1}{4}-18\pi KH_0r^2(\zeta_0+\zeta_1H_0)+\frac{6\pi KH_0r^3(\zeta_0+\zeta_1H_0)}{r-r_0}+\frac{144\pi^2K^2H_0^2r^5(\zeta_0+\zeta_1H_0)^2}{(r-r_0)(K-1)}. \label{18}
\end{equation}
One can find that $\alpha$ is a function of the radial coordinate $r$, which is similar to our previous work \cite{15} but different from Zaslavskii's work \cite{39}, since we consider the non-ideal equation of state here.

The second possibility is $b(r)=\sqrt{r_0r}$, obviously, which also satisfies the flare-out conditions. By using the same step as the first one, we get
\begin{equation}
\frac{U'}{U}=\frac{(1-\frac{K}{2})(r_0r)^{\frac{1}{2}}-24\pi H_0(\zeta_0+\zeta_1H_0)r^3}{r[r-(r_0r)^\frac{1}{2}]}. \label{19}
\end{equation}
Solving this equation, we obtain
\begin{equation}
U(r)=e^{\frac{1}{2}\{(K-2)\ln(r)-48H_0\pi(\zeta_0+\zeta_1H_0)(r+2\sqrt{r_0r})+[K-2+48r_0H_0\pi(\zeta_0+\zeta_1H_0)][2\arctan(\sqrt{\frac{r}{r_0}})-\ln(r-r_0)]\}}. \label{20}
\end{equation}
Hence, the solution is
\begin{equation}
ds^2=-e^{\frac{1}{2}\{(K-2)\ln(r)-48H_0\pi(\zeta_0+\zeta_1H_0)(r+2\sqrt{r_0r})+[K-2+48r_0H_0\pi(\zeta_0+\zeta_1H_0)][2\arctan(\sqrt{\frac{r}{r_0}})-\ln(r-r_0)]\}}dt^2+\frac{dr^2}{(1-\sqrt{\frac{r_0}{r}})}+r^2d\Omega^2. \label{21}
\end{equation}
Based on our previous work [14,15], we can also insert the redshift function $\Phi=C$ (C is a constant) into Eq. (\ref{9}) by hand
\begin{equation}
\frac{Kb'}{r^2}-b+24\pi H_0(\zeta_0+\zeta_1H_0)r^3=0. \label{22}
\end{equation}
It follows that
\begin{equation}
b(r)=C_1e^{\frac{r^3}{3K}}+24H_0\pi(3K+r^3)(\zeta_0+\zeta_1H_0), \label{23}
\end{equation}
where $C_1$ can be obtained through using the condition $b(r_0)=r_0$
\begin{equation}
C_1=[r_0-24H_0\pi(3K+r_0^3)(\zeta_0+\zeta_1H_0)]e^{-\frac{r_0^3}{3K}}. \label{24}
\end{equation}
So we can get
\begin{equation}
b(r)=e^{\frac{r^3-r_0^3}{3K}}[r_0-24H_0\pi(3K+r_0^3)(\zeta_0+\zeta_1H_0)]+24H_0\pi(3K+r^3)(\zeta_0+\zeta_1H_0), \label{25}
\end{equation}
and the solution becomes
\begin{equation}
ds^2=-e^{2C}dt^2+\frac{dr^2}{1-\frac{e^{\frac{r^3-r_0^3}{3K}}[r_0-24H_0\pi(3k+r_0^3)(\zeta_0+\zeta_1H_0)]+24H_0\pi(3K+r^3)(\zeta_0+\zeta_1H_0)}{r}}+r^2d\Omega^2. \label{26}
\end{equation}
\subsubsection{power law model: $\zeta=\eta H_0^\delta$}
In this model, the radial pressure that contains the bulk viscosity term can be expressed as follows
\begin{equation}
p_r=-K\rho-3\eta H_0^{1+\delta}, \label{27}
\end{equation}
where $\eta$ and $\delta$ are two parameters characterizing viscosity. Similarly, we just consider two cases $b(r)=r_0+\frac{1}{K}(r-r_0)$ and $\Phi=C$. For the first case, we can obtain
\begin{equation}
U(r)=\frac{C}{r}e^{-12K\pi \eta H_0^{1+\delta}}[r(2r_0+r)+2r_0^2ln(r-r_0)]. \label{28}
\end{equation}
Hence, the metric is
\begin{equation}
ds^2=-\frac{C}{r}e^{-12K\pi \eta H_0^{1+\delta}}[r(2r_0+r)+2r_0^2ln(r-r_0)]dt^2+\frac{dr^2}{(1-\frac{1}{K})(r-r_0)}+r^2d\Omega^2, \label{29}
\end{equation}
at the same time, we still suppose the anisotropic pressure is $p_\perp=\alpha\rho$, it follows that
\begin{equation}
\alpha(r)=\frac{K-1}{4}-18\pi K \eta H_0^{1+\delta}r^2+\frac{6\pi K\eta H_0^{1+\delta}r^3}{r-r_0}+\frac{144\pi^2\eta^2K^2H_0^{2(1+\delta)}r^5}{(r-r_0)(K-1)}. \label{30}
\end{equation}
For the second case, inserting $\Phi=C$ and Eq. (\ref{13}) into Eq. (\ref{9}) by hand, we get
\begin{equation}
\frac{Kb'}{r^2}-b+24\pi\eta H_0^{1+\delta} r^3=0. \label{31}
\end{equation}
It follows that
\begin{equation}
b(r)=C_1e^{\frac{r^3}{3K}}+24\pi\eta H_0^{1+\delta}(3K+r^3), \label{32}
\end{equation}
\begin{equation}
C_1=[r_0-24\pi\eta H_0^{1+\delta}(3K+r_0^3)]e^{-\frac{r_0^3}{3K}}. \label{33}
\end{equation}
Substituting Eq. (\ref{33}) into Eq. (\ref{32}), we can obtain
\begin{equation}
b(r)=e^{\frac{r^3-r_0^3}{3K}}[r_0-24\pi\eta H_0^{1+\delta}(3K+r_0^3)]+24\pi\eta H_0^{1+\delta}(3K+r^3), \label{34}
\end{equation}
so the solution is
\begin{equation}
ds^2=-e^{2C}dt^2+\frac{dr^2}{1-\frac{e^{\frac{r^3-r_0^3}{3K}}[r_0-24\pi\eta H_0^{1+\delta}(3K+r_0^3)]+24\pi\eta H_0^{1+\delta}(3K+r^3)}{r}}+r^2d\Omega^2. \label{35}
\end{equation}
\subsection{bulk viscosity as a function of present temperature: $\zeta=\zeta(T_0)$}
In paper \cite{a40}, the authors has extended the concept of temperature-dependent viscosity from classical statistical physics to observational cosmology, and take into account an interesting bulk viscosity model
\begin{equation}
\zeta=\xi T_0^{\epsilon}, \label{36}
\end{equation}
where $\xi$ and $\epsilon$ are two parameters characterizing viscosity. Furthermore, we are of interest to study the static and spherically symmetric wormhole solutions ar present time in this cosmological model, and the radial pressure of this model can be written as
\begin{equation}
p_r=-K\rho-3H_0\xi T_0^{\epsilon}. \label{37}
\end{equation}
For this model, similarly, we investigate the first case through inserting the shape function $b(r)=r_0+\frac{1}{K}(r-r_0)$ into Eq. (\ref{9}) by hand, we can get
\begin{equation}
\frac{U'}{U}=-\frac{1}{r}-\frac{24\pi\xi H_0T_0^{\epsilon}r^2}{(r-r_0)(1-\frac{1}{K})}. \label{38}
\end{equation}
where $T_0$ and $H_0$ denote the cosmological temperature at present time and present-day cosmic expansion velocity, respectively, and it follows that
\begin{equation}
U(r)=\frac{C}{r}e^{-12K\pi\xi H_0T_0^{\epsilon}}[r(2r_0+r)+2r_0^2ln(r-r_0)]. \label{39}
\end{equation}
Therefore, we can get the metric
\begin{equation}
ds^2=-\frac{C}{r}e^{-12K\pi\xi H_0T_0^{\epsilon}}[r(2r_0+r)+2r_0^2ln(r-r_0)]dt^2+\frac{dr^2}{(1-\frac{1}{K})(r-r_0)}+r^2d\Omega^2. \label{40}
\end{equation}
Then we continue considering the simplest anisotropic pressure $p_\perp=\alpha\rho$. After substituting Eq. (\ref{11}), Eq. (\ref{14}) and Eqs. (\ref{37}-\ref{38}) into Eq. (\ref{10}), we have
\begin{equation}
\alpha(r)=\frac{K-1}{4}-18K\pi\xi H_0T_0^{\epsilon}r^2+\frac{6K\pi\xi H_0T_0^{\epsilon}r^3}{r-r_0}+\frac{144\pi^2\xi^2K^2H_0^2T_0^{2\epsilon}r^5}{{(r-r_0)(K-1)}}. \label{41}
\end{equation}
Another case is still $\Phi=C$, we can easily get
\begin{equation}
b(r)=C_1e^{\frac{r^3}{3K}}+24H_0\xi\pi T_0^\epsilon(3K+r^3), \label{42}
\end{equation}
\begin{equation}
C_1=[r_0-24H_0\xi\pi T_0^\epsilon(3K+r_0^3)]e^{-\frac{r_0^3}{3K}}. \label{43}
\end{equation}
Thus, it follows that
\begin{equation}
b(r)=e^{\frac{r^3-r_0^3}{3K}}[r_0-24H_0\xi\pi T_0^\epsilon(3K+r_0^3)]+24H_0\xi\pi T_0^\epsilon(3K+r^3), \label{44}
\end{equation}
so the solution is
\begin{equation}
ds^2=-e^{2C}dt^2+\frac{dr^2}{1-\frac{e^{\frac{r^3-r_0^3}{3K}}[r_0-24H_0\xi\pi T_0^\epsilon(3K+r_0^3)]+24H_0\xi\pi T_0^\epsilon(3K+r^3)}{r}}+r^2d\Omega^2. \label{45}
\end{equation}
\subsection{bulk viscosity as a function of energy density: $\zeta=\zeta(\rho_0)$}
In this situation, we just consider the power law model, i.e., the bulk viscosity coefficient can be expressed as
\begin{equation}
\zeta=\beta\rho_0^\gamma, \label{46}
\end{equation}
where $\beta$ and $\gamma$ are two parameters characterizing viscosity, $\beta>0$ ensures a positive entropy at the request of the second law of thermaldynamics. So the radial pressure after substituting Eq. (\ref{8}) into Eq. (\ref{46}) becomes
\begin{equation}
p_r=-\frac{1}{8\pi r^2}-3H_0\beta(8\pi Kr^2)^{-\gamma}. \label{47}
\end{equation}
In the first place, we still start from the shape function $b(r)=r_0+\frac{1}{K}(r-r_0)$,
then Eq. (\ref{9}) is
\begin{equation}
\frac{U'}{U}=-\frac{1}{r}-\frac{24\pi\beta H_0(8\pi Kr^2)^{-\gamma} r^2}{(r-r_0)(1-\frac{1}{K})}, \label{48}
\end{equation}
 we obtain by using Eq. (\ref{47}) in Eq. (\ref{48})
\begin{equation}
U(r)=\frac{C}{r}e^{-12K\pi\beta H_0(8\pi Kr^2)^{-\gamma}}[r(2r_0+r)+2r_0^2ln(r-r_0)]. \label{49}
\end{equation}
Hence, the metric is
\begin{equation}
ds^2=-\frac{C}{r}e^{-12K\pi\beta H_0(8\pi Kr^2)^{-\gamma}}[r(2r_0+r)+2r_0^2ln(r-r_0)]dt^2+\frac{dr^2}{(1-\frac{1}{K})(r-r_0)}+r^2d\Omega^2. \label{50}
\end{equation}
It is worth noting that the energy density $\rho$ still represents $(8K\pi r^2)^{-1}$. Similarly, here we just consider the anisotropic pressure as $p_\perp=\alpha\rho$ and we have through some calculations
\begin{equation}
\alpha(r)=\frac{K-1}{4}-\frac{3}{2}\gamma\beta H_0(8\pi K)^{2-\gamma}r^{5-2\gamma}-\frac{9}{4}H_0\beta(8\pi Kr^2)^{1-\gamma}+\frac{6\pi K\beta(8\pi Kr^2)^{-\gamma} r^3}{r-r_0}+\frac{18K\beta^2H_0^2(8\pi K)^{1-2\gamma}r^{5-4\gamma}}{(r-r_0)(K-1)}. \label{51}
\end{equation}
In the second place, through assuming the redshift function $\Phi=C$ and inserting it and Eq. (\ref{8}) into Eq. (\ref{9}) we can have
\begin{equation}
\frac{Kb'}{r^2}-b+24\pi H_0\beta(\frac{b'}{8\pi r^2})^\gamma r^3=0. \label{52}
\end{equation}
Here we just concentrate on the case of $\gamma$=1, then Eq. (\ref{50}) can be analytically solved as follows
\begin{equation}
b(r)=C_1(K+3H_0r^3\beta)^{\frac{1}{9H_0\beta}}. \label{53}
\end{equation}
Use the condition $b(r_0)=r_0$ once again,
\begin{equation}
C_1=r_0(K+3H_0r_0^3\beta)^{-\frac{1}{9H_0\beta}}, \label{54}
\end{equation}
After substitution of Eq. (\ref{54}) into Eq. (\ref{53}), we get
\begin{equation}
b(r)=r_0(\frac{K+3H_0r^3\beta}{K+3H_0r_0^3\beta})^{\frac{1}{9H_0\beta}}. \label{55}
\end{equation}
So the metric is
\begin{equation}
ds^2=-e^{2C}dt^2+\frac{dr^2}{1-\frac{r_0}{r}(\frac{K+3H_0r^3\beta}{K+3H_0r_0^3\beta})^{\frac{1}{9H_0\beta}}}+r^2d\Omega^2. \label{56}
\end{equation}
It is worth noting that the anisotropic solutions for the case $b(r)=r_0+\frac{1}{K}(r-r_0)$ in three different physical considerations from Eq. (\ref{18}), Eq. (\ref{30}), Eq. (\ref{41}) and  Eq. (\ref{51}) have substantial high degeneracy with each other. To be more precise, the first term $\frac{K-1}{4}$ in these equations corresponds to the anisotropic solution of the ideal cosmic fluid ($p=-K\rho$). In principle, one can obtain the same expression of anisotropic solution by choosing appropriately the parameters $\zeta_0$, $\zeta_1$, $H_0$, $\eta$, $\delta$, $\xi$, $\epsilon$, $T_0$, $\gamma$ and $\beta$. Thus, to some extent, these three classes of viscous models have very high degeneracy with each other and deviate slightly from the ideal fluid consequence.
\section{Traversabilities Analysis}
In wormhole physics, the most attractive thing may be to analyze the traversabilities including traversal time and traversal velocity for a human being to journey through the wormhole. To perform the traversabilities analysis, we consider the case of the constant redshift function for the linear model $\zeta=\zeta_0+\zeta_1H$, therefore, the corresponding shape function is shown in Eq. (\ref{25}). Without the loss of generality, we set $\zeta_1=0$, thus, the shape function can be rewritten as
\begin{equation}
b(r)=e^{\frac{r^3-r_0^3}{3K}}[r_0-24H_0\pi\zeta_0(3K+r_0^3)]+24H_0\pi\zeta_0(3K+r^3), \label{57}
\end{equation}
It is easy to be seen that this solution is only traversable but not asymptotically flat, since $\Phi$ is finite and $b/r\nrightarrow0$ when $r\rightarrow\infty$. Hence, the dimensions of this wormhole may be very finite.

As noted in our previous works \cite{40,41}, three conditions must be satisfied for a human being in the spaceship who intends to journey through the wormhole.

$\star$ The acceleration felt by the travelers should not exceed 1 Earth's gravitational acceleration $g$:
\begin{equation}
\left|(1-\frac{b}{r})^{\frac{1}{2}}(\tau e^\Phi)'e^{-\Phi}\right|\leq g,
\end{equation}
where $\tau=(1-v^2)^{-1/2}$.
 
$\star$ The tidal acceleration should not exceed 1 Earth's gravitational acceleration $g$:
\begin{equation}
\left|\iota^1\right|\left|(1-\frac{b}{r})[\Phi''+(\Phi')^2-\frac{b'r-b}{2r(r-b)}\Phi']\right|\leq g,
\end{equation}
\begin{equation}
\left|\iota^2\right|\left|\frac{\tau^2}{2r^2}[v^2(b'-\frac{b}{r})+2(r-b)\Phi']\right|\leq g,
\end{equation}
where $v$ is the traveler's velocity and  $\left|\iota^i\right|\approx2 m$ ($i=1,2$) is the size of the traveler, namely the distance between two arbitrary parts of the traveler's body \cite{8}.

$\star$ The traverse time measured by the travelers and the observers who keep at rest at the space station should satisfy the following quantitative relationships ( see more mathematical details in \cite{8,42}):
\begin{equation}
\Delta T=\int^{+l_2}_{-l_1}\frac{dl}{v\tau},
\end{equation}
\begin{equation}
\Delta t=\int^{+l_2}_{-l_1}\frac{dl}{ve^\Phi},
\end{equation}
where $dl=(1-\frac{b}{r})^{-1/2}dr$ is the proper radial distance, and we consider that the space stations are located at a radius $r=a$, at $-l_1$ and $l_2$, respectively.

One can easily find that inequalities (58) and (59) are naturally satisfied in the case of a constant redshift function. For a constant non-relativistic traversal velocity ($\tau\approx1$), replacing Eq. (\ref{57}) in Eq. (60), we obtain the traversal velocity of this viscous model as follows
\begin{equation}
v\leq r_0\sqrt{\frac{Kg}{r_0^3-24H_0\pi\zeta_0r_0^5-K}}, \label{58}
\end{equation}
Subsequently, assuming the throat radius of the wormhole $r_0=10$ m, the dark energy equation of state parameter $K=2$, the present-day Hubble parameter $H_0=73.24$ $kms^{-1}Mpc^{-1}$ and the bulk viscosity coefficient $\zeta_0=10^{-6}$, we obtain the traversal velocity $v\approx2.10$ m/s. Furthermore, we find that the value of traversal velocity decreases for decreasing bulk viscosity, and approaches $v\approx1.40$ m/s. In the meanwhile, we also find that the traversal velocity for a traveler depends on not only the wormhole geometry ($r_0$) but also the effects of cosmological background evolution ($K$, $\zeta_0$ and $H_0$). Note that, in the above, we actually compute out the traversal velocity of the viscous $\omega$CDM model with a constant bulk viscosity coefficient.

\section{Energy Conditions}
Generally speaking, there are two important kinds of energy condition in classical general relativity, namely, the point-wise energy conditions and averaged energy conditions. The point-wise energy conditions usually include the above-mentioned NEC, the weak energy condition (WEC), the strong energy condition (SEC) and the dominant energy condition (DEC). More precisely, the corresponding mathematical descriptions for different energy conditions in a spatially flat Friedmann-Robertson-Walker (FRW) universe are expressed as

$\star$ NEC $\Leftrightarrow \rho+p\geqslant0$,

$\star$ WEC $\Leftrightarrow \rho+p\geqslant0\quad$ and $\quad\rho\geqslant0$,

$\star$ SEC $\Leftrightarrow \rho+p\geqslant0\quad$ and $\quad\rho+3p\geqslant0$,

$\star$ DEC $\Leftrightarrow \rho\pm p\geqslant0\quad$ and $\quad\rho\geqslant0$.

It is worth noting that, mathematically, the NEC is the necessary condition of the other three energy conditions. In what follows, we would like to investigate the behaviors of different energy conditions in astrophysical scales and explore the effects of the bulk viscosity on the energy conditions. Here we still consider the above-mentioned case $\Phi=C$ and $p=-K\rho-3\zeta_0H$. Subsequently, using Eq. (\ref{8}) and Eq. (\ref{57}), one can easily obtain the NEC as
\begin{equation}
p+\rho=\frac{1-K}{8\pi}\{\frac{e^{\frac{r^3-r_0^3}{3K}}}{K}[r_0-24H_0\pi\zeta_0(3K+r_0^3)]+72H_0\pi\zeta_0\}-3\zeta_0H_0\geqslant0. \label{59}
\end{equation}

\begin{figure}
\centering
\includegraphics[scale=0.38]{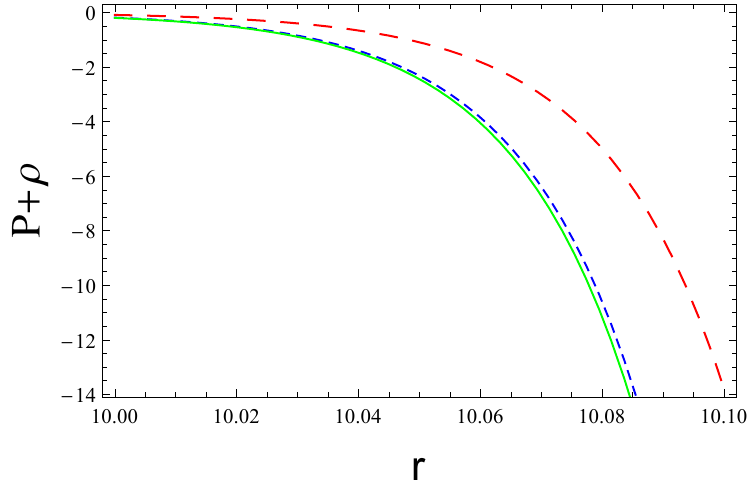}
\includegraphics[scale=0.38]{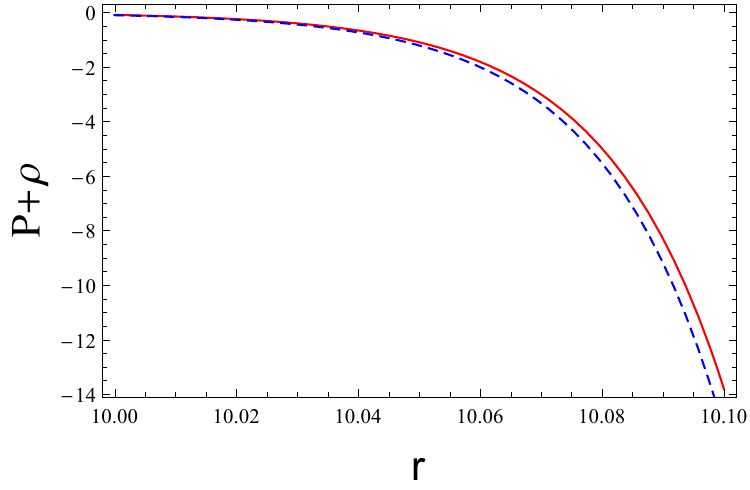}
\includegraphics[scale=0.38]{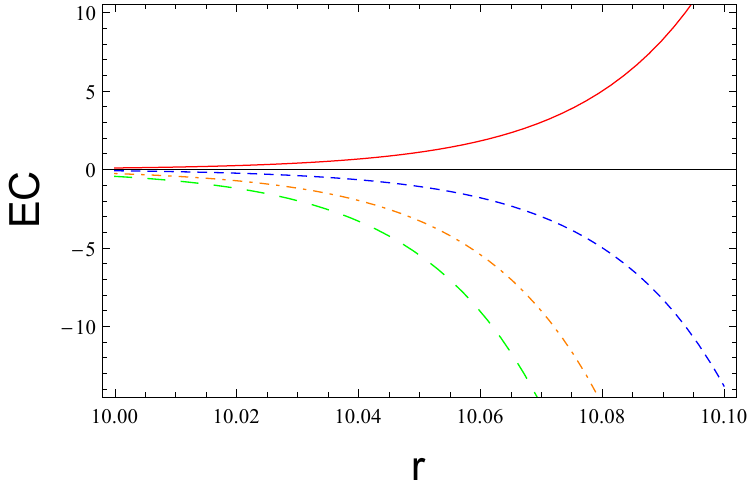}
\caption{The left panel represents the relation between $p+\rho$ and $r$: for given throat radius of the wormhole $r_0=10$, dark energy equation of state parameter $K=2$ and Hubble parameter $H_0=73.24$, the red (long-dashed) line corresponds to the case $\zeta_0=10^{-6}$, the blue (short-dashed) line $\zeta_0=10^{-7}$ and the green (solid) line $\zeta_0=10^{-8}$. The medium panel represents the relation between $p+\rho$ and $r$: for given throat radius of the wormhole $r_0=10$, dark energy equation of state parameter $K=2$ and the bulk viscosity coefficient $\zeta_0=10^{-6}$, the red (solid) line corresponds to $H_0=73.24$ and the blue (dashed) line $H_0=66.93$ (note that we just take the best fitting value of the Hubble parameter). The right panel represents the relations between different energy conditions and $r$: for given parameters $\zeta_0=10^{-6}$, $K=2$, $r_0=10$ and $H_0=73.24$, the red (solid) line corresponds to $\rho$, the blue (short-dashed) line $\rho+3p$, the orange line $p+\rho$, the green (long-dashed) line $p-\rho$ and the horizontal line zero.}\label{ff1}
\end{figure}

Furthermore, in the left panel of Fig. \ref{ff1}, we find that the NEC will be violated more clearly when the bulk viscosity decreases, for given throat radius of the wormhole $r_0$, dark energy equation of state parameter $K$ and Hubble parameter $H$.
In the medium panel of Fig. \ref{ff1}, using the recent Planck's result and A. Riess et al.'s result \cite{3,43}, we also show that the NEC is violated more clearly when   the value of the Hubble parameter decreases, for given bulk viscosity coefficient $\zeta_0$, throat radius of the wormhole $r_0$ and dark energy equation of state parameter $K$. This means that, the more clearly the NEC is violated, the smaller the cosmic expansion velocity is. In the right panel of Fig. \ref{ff1}, we also exhibit the relations between different energy conditions and the radial coordinate $r$, and find that the NEC, DEC and SEC are always violated when $r>r_0$, except for $\rho\geqslant0$ of the WEC is satisfied all the time for given parameters $\zeta_0$, $K$, $r_0$ and $H_0$. Physically, the non-violation of the WEC indicates that the dark energy fluid in the viscous $\omega$CDM model has a non-diverging effect on congruences of null geodesics. However, since the NEC in this case is violated, all the point-wise energy conditions are always violated when $r>r_0$.
\begin{figure}
\centering
\includegraphics[scale=0.38]{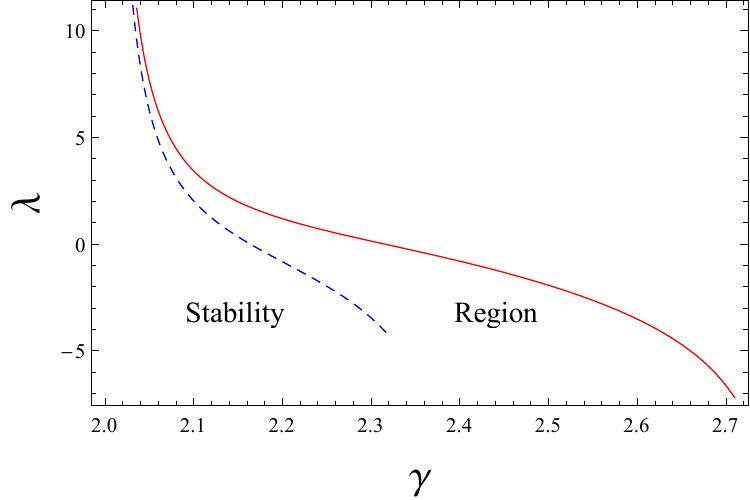}
\includegraphics[scale=0.38]{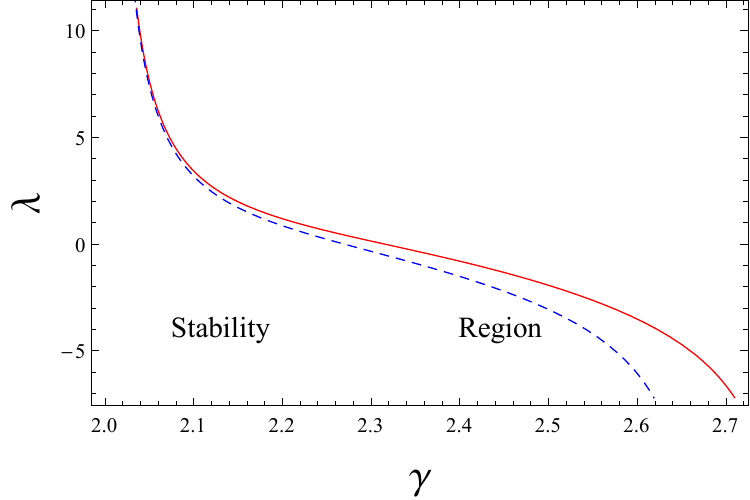}
\includegraphics[scale=0.38]{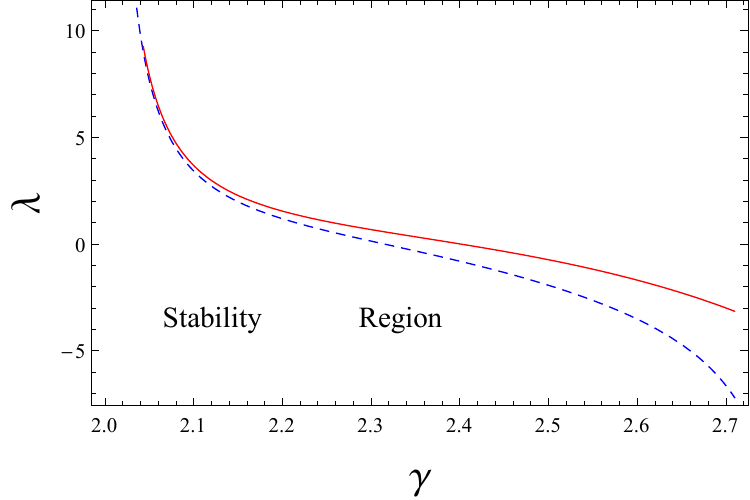}
\caption{Plots for a positive surface energy density, i.e., $b(a_0)<2M$. In the left panel, we set the parameters $K=2$, $H_0=73.24$ and $\zeta_0=10^{-6}$ for both cases. The red (solid) and blue (dashed) lines represent $r_0/M=0.1$ and $r_0/M=0.25$, respectively. In the medium panel, we set the parameters $K=2$, $H_0=73.24$ and $r_0/M=0.1$ for both cases. The red (solid) and blue (dashed) lines represent $\zeta_0=10^{-6}$ and $\zeta_0=10^{-10}$, respectively. In the right panel, we set the parameters $K=2$, $r_0/M=0.1$ and $\zeta_0=10^{-6}$ for both cases. The red (solid) and blue (dashed) lines represent $H_0=150$ and $H_0=73.24$, respectively. The stability regions are shown beneath these curves.}\label{ff2}
\end{figure}

\section{Stability Analysis}
In this section, we investigate the effects of the bulk viscosity on stability of the wormhole spacetime configuration. Concretely, we still take into account the case of $\Phi=C$ and $p=-K\rho-3\zeta_0H_0$. Furthermore, we conclude the standard potential method in \cite{42} as follows:

$\star$ Use the Darmois-Israel formalism to derive the so-called Lanczos equation, and match an exterior Schwarzschild geometry into the interior traversable wormhole geometry.

$\star$ Work out the nontrivial components of the extrinsic curvature, the surface energy density $\sigma$ and tangential pressure $P$, the defined physical quantity $\Xi$ and the radial derivative of the surface energy density $\sigma'$. Note that these quantities are all the functions of the junction radius $a$.

$\star$ Utilize the equation of motion and linearize around a stable solution situated at $a_0$ to derive the equilibrium relation and the stability condition.

$\star$ Define a new quantity $\lambda$ to characterize the stable equilibrium and combine the master equation and the aforementioned stability condition to reexpress the stability condition (see Eqs. (41-42) in \cite{42}). Note that the new stable condition shall be discussed separately in the cases of a positive and a negative surface energy density.

$\star$ Substitute the redshift function $\Phi(a)$ and the shape function $b(a)$ of a given cosmological scenario into the new stable condition, so that one can get the final stability expression. However, in general, these results shall be exhibited graphically since the expression of the parameter $\lambda$ is substantially cumbersome.

In what follows, we would like to discuss separately the cases of $b(a_0)<2M$ and $b(a_0)>2M$, i.e., the cases of a positive and a negative surface energy density.

For the case of $b(a_0)<2M$, we obtain the range of the matching radius as follows
\begin{equation}
2M<a_0<\{\frac{M}{12H_0\pi\zeta_0}-3K-3K\times P(-\frac{e^{\frac{M}{36H_0K\pi\zeta_0}(72H_0K\pi\zeta_0+24H_0\pi\zeta_0r_0^3-r_0)-\frac{r_0^3}{3K}-1}}{72H_0K\pi\zeta_0})\}^{\frac{1}{3}}, \label{60}
\end{equation}
where $P(x)$ represents the Lambert-W function, also called product logarithm function $ProductLog(x)$, which denotes two single-value branches of the inverse function of the function $f(x)=xe^x$. More specifically, these two single-value branches lie in the range $(-\infty,-1/e)$ and $(-1/e, \infty)$, respectively. For example, $P(5)=1.3267$, $P(500)=4.6728$, etc.

\begin{figure}
\centering
\includegraphics[scale=0.38]{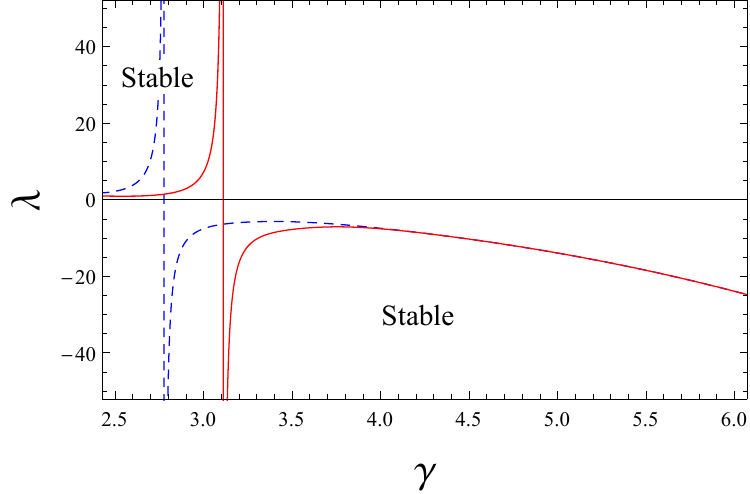}
\includegraphics[scale=0.38]{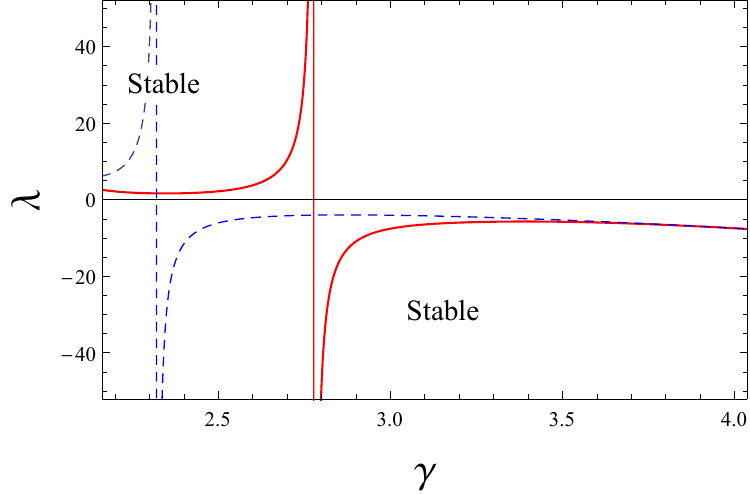}
\includegraphics[scale=0.38]{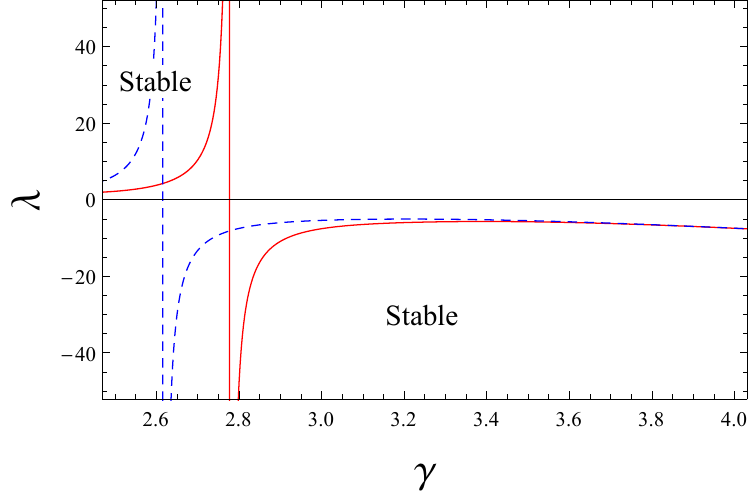}
\caption{Plots for a negative surface energy density, i.e., $b(a_0)>2M$. In the left panel, we set the parameters $K=2$, $H_0=73.24$ and $\zeta_0=10^{-6}$ for both cases. The red (solid) and blue (dashed) lines represent $r_0/M=0.03$ and $r_0/M=0.01$, respectively. In the medium panel, we set the parameters $K=2$, $H_0=73.24$ and $r_0/M=0.01$ for both cases. The red (solid) and blue (dashed) lines represent $\zeta_0=10^{-6}$ and $\zeta_0=10^{-5}$, respectively. In the right panel, we set the parameters $K=2$, $r_0/M=0.01$ and $\zeta_0=10^{-6}$ for both cases. The red (solid) and blue (dashed) lines represent $H_0=73.24$ and $H_0=150$, respectively. In the upper half panel, the stability regions are shown above the curves. In the lower half panel, the stability regions are shown beneath the curves.}\label{ff3}
\end{figure}

By fixing the parameters $K=2$, $H_0=73.24$ and $\zeta_0=10^{-6}$, we consider the following cases: $r_0/M=0.25$, in order that $2<a_0/M<2.35$; for $r_0/M=0.1$, we obtain $2<a_0/M<2.71$. One can easily find that the range of the junction radius shall decrease for increasing values of $r_0/M$. Subsequently, we also exhibit the the respective stability regions in the left panel of Fig. \ref{ff2}, and find that, for increasing values of $r_0/M$, regardless of the decreasing range $a_0$, the values of the parameter $\lambda$ are further restricted. Subsequently, we are very interested in investigating the viscous effects on stability of the wormhole configuration. By fixing the parameters $K=2$, $H_0=73.24$ and $r_0/M=0.1$, we take into account the following cases: $\zeta_0=10^{-6}$, so that $2<a_0/M<2.71$; for $\zeta_0=10^{-10}$, we obtain $2<a_0/M<2.619$. It is easy to be seen that the range of the junction radius shall increase for increasing values of $\zeta_0$. In the meanwhile, for decreasing values of $\zeta_0$, regardless of the decreasing range $a_0$, the values of the parameter $\lambda$ are further restricted (see the medium panel of Fig. \ref{ff2}). In addition, we also explore about the effects of cosmic background evolution on stability of the wormhole configuration. By fixing the parameters $K=2$, $r_0/M=0.1$ and $\zeta_0=10^{-6}$, we consider the following cases: $H_0=73.24$, in order that $2<a_0/M<2.71$; for $H_0=150$, we obtain $2<a_0/M<2.87$. It is obvious that the range of the junction radius increases for increasing values of $H_0$. From the right panel of Fig. \ref{ff2}, we can find that, for decreasing values of $H_0$, regardless of the increasing range $a_0$, the values of the parameter $\lambda$ are further restricted.

For the case of $b(a_0)>2M$, we obtain the range of the matching radius as
\begin{equation}
a_0>\{\frac{M}{12H_0\pi\zeta_0}-3K-3K\times  P(-\frac{e^{\frac{M}{36H_0K\pi\zeta_0}(72H_0K\pi\zeta_0+24H_0\pi\zeta_0r_0^3-r_0)-\frac{r_0^3}{3K}-1}}{72H_0K\pi\zeta_0})\}^{\frac{1}{3}}. \label{61}
\end{equation}

For $r_0<2M$, using the same method described in \cite{42}, we just consider the case of $K=2$. By fixing the parameters $H_0=73.24$ and $\zeta_0=10^{-6}$, we take into account the following cases: $r_0/M=0.01$, so that $a_0/M>2.51$; for $r_0/M=0.03$, we obtain $a_0/M>2.45$. The asymptotes $\sigma'|_R=0$ for these cases situate at $R/M\thickapprox2.77$ and $R/M\thickapprox3.12$, respectively. The corresponding stability regions are exhibited in the left panel of Fig. \ref{ff3}. For increasing values of $r_0/M$, the range of the junction radius increases, the values of the parameter $\lambda$ are less restricted and one may conclude that the total stability region increase. In the medium panel of Fig. \ref{ff3}, we investigate the viscous effects on stability of the wormhole configuration. Subsequently, we find that, for decreasing values of $\zeta_0$, the range of the junction radius increases, the values of the parameter $\lambda$ are less restricted and one may still conclude that the total stability region increase. Furthermore, it is very constructive to study the the effects of cosmic background evolution on stability of the wormhole configuration. From the right panel of Fig. \ref{ff3}, we find that, for increasing values of $H_0$, the range of the junction radius decreases, the values of the parameter $\lambda$ are less restricted and one may conclude that the total stability region increase.

For $r_0>2M$, we also demonstrate that the values of $\lambda$ are always negative. By increasing $r_0/M$, decreasing $\zeta_0$ and increasing $H_0$, the values of the parameter $\lambda$ are less restricted, one may conclude that the total stability region increases, and the range of the junction radius increases, increases and decreases, respectively.

\section{Discussions and conclusions}
Since the elegant discovery that our universe is undergoing the phase of accelerated expansion, more recently, theorists have gradually paid more and more attention to the exotic spacetime configurations, especially, the renewed field---wormholes. In the literature, all the authors almost construct wormholes by only using the linear equation of state, namely, assuming that the universe is filled with a perfect fluid. Nonetheless, the dynamical behaviors of the true universe should be physically dominated by a non-ideal cosmic fluid. Therefore, we are very interested in exploring wormholes in the bulk viscosity cosmology.

In this paper, at first, we make a review about the mathematical descriptions of the wormholes. Subsequently, three classes of viscous models are considered, i.e., bulk viscosity as a function of Hubble parameter $H$, temperature $T$ and dark energy density $\rho$, respectively. In general, the larger the cosmic expansion velocity $H$, the smaller the bulk viscosity of the universe is; the larger the cosmic temperature is, the smaller the bulk viscosity of the universe is; the larger the dark energy density, the larger the the bulk viscosity of the universe is. 
For every model, we solve the Einstein field equations by both inserting the shape function $b(r)=r_0+\frac{1}{K}(r-r_0)$ and the redshift function $\Phi=C$. Since Eq. (\ref{18}), Eq. (\ref{30}), Eq. (\ref{41}) and  Eq. (\ref{51}) can be transformed to each other mathematically by choosing appropriately the parameters $\zeta_0$, $\zeta_1$, $H_0$, $\eta$, $\delta$, $\xi$, $\epsilon$, $T_0$, $\gamma$ and $\beta$. Thus, to some extent, these three classes of viscous models have very high degeneracy with each other and deviate slightly from the ideal fluid consequence ($p=-K\rho$). In what follows, without the loss of generality, to perform the traversabilities, energy conditions and stability for the wormhole solution, we consider the case of of $\Phi=C$ and $p=-K\rho-3\zeta_0H$, namely, the viscous $\omega$CDM with a constant bulk viscosity coefficient. We find that the value of traversal velocity decreases for decreasing bulk viscosity, and the traversal velocity for a traveler depends on not only the wormhole geometry ($r_0$) but also the effects of cosmological background evolution ($K$, $\zeta_0$ and $H_0$). Subsequently, we find that the NEC will be violated more clearly when the bulk viscosity decreases; the more clearly the NEC is violated, the smaller the cosmic expansion velocity is; the NEC, DEC and SEC are always violated when $r>r_0$, except for $\rho\geqslant0$ of the WEC is satisfied all the time. Physically, the non-violation of the WEC indicates that the dark energy fluid in the viscous $\omega$CDM model has a non-diverging effect on congruences of null geodesics. However, since the NEC in this case is violated, all the point-wise energy conditions are always violated when $r>r_0$ (see Fig. \ref{ff1}). By implementing the stability analysis, we find that for the case of a positive surface energy density, the range of the junction radius decreases and the values of the parameter $\lambda$ are further restricted when $r_0/M$ increases, $\zeta_0$ decreases and $H_0$ decreases, respectively (see Fig. \ref{ff2}); for the case of a negative surface energy density, by increasing $r_0/M$, decreasing $\zeta_0$ and increasing $H_0$, the values of the parameter $\lambda$ are less restricted, one may conclude that the total stability region increases, and the range of the junction radius increases, increases and decreases, respectively (see Fig. \ref{ff3}).

Note that the condition $0<\lambda\leqslant1$ in paper \cite{42}, which corresponds to the requirement that the speed of sound should not exceed the speed of light, should be relaxed in the presence of the exotic matter since we still know little about the nature and cosmological origin of the dark energy.

In the future, we expect that more and more highly accurate data can provide newer knowledge about various kinds of celestial bodies and cosmological background evolution for us.

\section{acknowledgements}
Deng Wang warmly thanks Liang Gao, Jie Wang and Qi Guo for beneficial feedbacks on astrophysics and cosmology. The authors acknowledge partial support from the National Science Foundation of China. This work is supported by National Nature Science Foundation of China under Grants No.11988101 and No.11851301.

\end{document}